\documentstyle[sprocl]{article}

\def\Journal#1#2#3#4{{#1} {\bf #2}, #3 (#4)}

\def\NPB{{\em Nucl. Phys.} B}
\def\PLB{{\em Phys. Lett.}  B}

\def\PRD{{\em Phys. Rev.} D}

\def\YF{\em Yad. Fiz.}

\def\bea{\begin{eqnarray}}
\def\eea{\end{eqnarray}}
\def\beq{\begin{equation}}
\def\eeq{\end{equation}}

\def\oq{\omega(q)}

\def\eq{\eta (q)}
\def\ea{\eta (q_{1})}
\def\eb{\eta (q_{2})}
\def\ec{\eta (q'_{1})}
\def\ed{\eta (q'_{2})}

\def\xq{\xi(q)}
\def\xa{\xi(q_1)}
\def\xb{\xi(q_2)}

\begin{document}

\title{Interaction of Two Reggeized Gluons: 2nd Order Corrections
and the Bootstrap}

\author{G.P. Vacca}

\address{II Institut f\"ur Theoretische Physik, Universit\"at Hamburg 
\\E-mail: vacca@bo.infn.it} 


\maketitle\abstracts{Our aim is to show how the reggeization of the gluon,
encoded in the bootstrap property of the BFKL kernel, permits to calculate the
interaction kernel in the octet colour channel in the forward and non forward
direction, for the quark contribution, starting from the known gluon
trajectory up to 2nd order. 
To this end an ansatz to solve the bootstrap equation is used. 
The obtained result, for what regard the quark contribution, 
has been verified by direct Feynman diagram calculations by Fadin et al..}
\centerline{
Talk at Internatinal Conference on Elastic and Diffractive Scattering}
\centerline{(VIIIth EDS Blois Workshop), Protvino, Russia, 
June 28 - July 2, 1999.}
\centerline{(to be published in the Proceeding of this Conference)}

\section{Introduction}
The BFKL equation and its impact in small $x$ physics have been widely
discussed subjects in the last years and recently the 2nd order corrections
were calculated for the forward pomeron equation \cite{bfklnla1,bfklnla2}.
Generalization of this result in the non forward direction would, of course,
be really interesting, not only in the singlet colour channel of the pomeron.
For the other physically important case, that of the odderon,
the 2nd order corrections to the non-forward interaction of two reggeized
gluons in the octet colour state is needed (as well as also the form of the
gluon triple interaction term).

This interaction is severely restricted by the so-called bootstrap relation
which guarantees that
production amplitudes with the gluon quantum number in their $t$
channels used for the construction of the absorptive part are indeed
given only by a single reggeized gluon exchange and do not contain
admixture from two or more reggeized gluon exchanges \cite{lip2,bart1}.
The bootstrap relation is known to be satisfied in the lowest order
in the coupling constant.
The form this relation takes in the 2nd order was recently discussed
by Fadin et al. \cite{fafio}

The bootstrap can be expressed as a relation which involve the reggeized
gluon trajectory, the interaction between two reggeized gluons and their
wave function.
We show here \cite{bravac} how, using the bootstrap in a form stronger than
in \cite{fafio} and an ansatz \cite{braun} to solve it, it is possible
to obtain the full second order corrections,
at least due to the quark contribution, for the non-forward gluon interaction
in the octet channel necessary for the odderon equation.
It can also be shown that we also automatically find
the corrections for the running of the coupling for the non-forward pomeron
equation \cite{bravac}.
\section{The Bootstrap relation}
We find convenient to work with an operator formalism to describe the
two gluon equation and the bootstrap relation.
Let the two gluons with momenta $q_{1,2}$ be described by a wave function
$\Psi(q_1,q_2)$. The total momentum $q=q_1+q_2$ will always be conserved,
so that in future $q_2=q-q_1$ and the dependence on $q$ and thus $q_2$
will be suppressed.
The scalar product of two wave functions will be taken as
$
\langle\Psi_1|\Psi_2\rangle\equiv\int d^{D-2}q_1
\Psi_1^*(q_1)\Psi_2(q_1),
$
where $D=4+2\epsilon$, $\epsilon\rightarrow 0$, is the dimension used for
the infrared regularization.

In this metric, as a function of the angular momentum $j$ and
up to  order $\alpha^3_s$, the absorptive part of the
scattering amplitude coming
from the two-reggeized-gluon exchange is given by
$
{\cal A}_j=\langle\Phi_p|G(E)|\Phi_t\rangle
$.

Here the  functions $\Phi_{p,t}$ are the so called impact factors
 which represent
the coupling of the external particles ($t$ from the target and $p$
from the projectile) to the two exchanged gluons and are of order
$\alpha_s=g^2/4\pi$. 

The Green function $G(E)$ with $E=1-j$ is defined as
$
  G(E)=(H-E)^{-1}
$
where $H$ is the two gluon (Hermithean) Hamiltonian. Its explicit
expression in the momentum representation is
\beq
\langle q_1|H|q'_1\rangle=-\delta^{D-2}(q_1-q'_1)(\omega(q_1)+
\omega(q_2))-V^{(R)}(q_1,q'_1)
\eeq
where $1+\omega(q_{1(2)})$ is the Regge trajectory of the first 
(second) gluon and $V^{(R)}$ represents the gluonic interaction
for the $t$-channel with the colour quantum number $R$.

In any colour channel the Green function $G(E)$ can be represented via
the orthonormalized solutions of the homogeneous Schr\"odinger equation
$
H\Psi_n=E_n\Psi_n
$.
Therefore, in terms of the eigenfunctions $\Psi_n$
the absorptive part ${\cal A}_j$ is
\beq
{\cal A}_j=
\sum_n\frac{\langle\Phi_p|\Psi_n\rangle\langle\Psi_n|\Phi_t
\rangle}{E_n-E}.
\label{Aexp}
\eeq

Let us consider the case with a colour quantum number of
the gluon ($R=g$) in the $t$ channel.
The BFKL equation is based on the idea
that physical amplitudes in this channel have the asymptotical behaviour
corresponding to an exchange of a single reggeized gluon
so that no contribution should come from the exchange of
two or more reggeized gluons (in the leading (LO) and next-to-leading (NLO)
orders).
This means that the absorptive part, as a function
$j$, should have a simple pole at
$j=1+\omega(q)$ and no other singularities.
We can therefore state the following properties valid in the {\bf LO and NLO}:

{\bf 1.} the spectrum of $H$ in the considered $t$ channel should contain
an eigenvalue $E_0=-\omega(q)$ with the corresponding eigenfunction
$\Psi_0$
\beq
(H+\omega(q))\Psi_0=0
\label{strboot1}
\eeq

{\bf 2.} the expression in (\ref{Aexp}) take contributions only from
this eigenfunctions.
This means that both impact factors $\Phi_{p,t}$ should be
orthogonal to all other eigenfunctions $\Psi_n$, $n>0$. Due to the
completeness of the whole set it means that both $\Phi_p$ and
$\Phi_t$ should coincide with $\Psi_0$, up to a normalization factor,
which will contain all the dependencies in the other variables
(factorization).

{\bf 3.} there is a relation for the impact factor.
In fact one considers the Mellin transform for the amplitude in (\ref{Aexp})
where only the contribution for $n=0$ is present in the sum. Taking into
account the proper signature factor and comparing to the standard form,
where the coupling $\Gamma$'s
(particle-particle-reggeon vertices) are factorized from the contribution
of the reggeized gluon, one obtains the ``third bootstrap condition''
\beq
\Gamma(q)=
q\left(\frac{q^2}{s_0}\right)^{\omega(q)/2}
\frac{\langle\Phi|\Psi_0\rangle}{\sqrt{|\sin \pi\omega(q)|}}.
\label{boot3}
\eeq
Considering a perturbative approach one may write
$\Psi_0=\chi+\Psi_0^{(1)}$, $\oq=\omega^{(1)}(q)+\omega^{(2)}(q)$ and
$H=H^{(1)}+H^{(2)}$ and obtain at the second order for the bootstrap eq.
(\ref{strboot1}) the form
\beq
(H^{(1)}+\omega^{(1)}(q))\Psi_0^{(1)}=
-(H^{(2)}+\omega^{(2)}(q))\chi
\eeq
This general relation may be projected onto the state $\chi$ (the first order
wave function) to obtain the relation
\beq
\langle\chi|H^{(2)}+\omega^{(2)}(q)|\chi\rangle=0
\label{weakboot1}
\eeq
which has been previously obtained \cite{fafio} from different arguments
together with the second order form of (\ref{boot3}).
We shall denote the relation (\ref{weakboot1}) as a weak bootstrap condition.
\section{Bootstrap at work}
The main idea consists in taking one particular solution of the bootstrap
relation, which describe all the inter-dependent quantities in terms of a
single function. Explicitely in the LO BFKL equation every
transverse momentum squares are substituted by a function of them which we
call $\eta$.
Explicitely taking:
\beq
\oq=-\int
d^{2}q_{1}\frac{\eq}{\ea\eb}
\label{traj}
\eeq
and
\beq
V^{(g)}(q_1,q'_{1})=\frac{1}{\sqrt{\ea\eb\ec\ed}}
\left(\frac{\ea\ed+\eb\ec}{\eta(k)}-\eq\right)
\label{pot}
\eeq
then Eq.(\ref{strboot1}) remains valid with a solution
\beq
\Psi_0=1/\sqrt{\ea\eb}
\label{state}
\eeq
Therefore the strategy to determine the interaction potential
for the two gluons in the octet state will be the following: \
\noindent {\it \underline {known gluon trajectory at NLO} $\to$ 
\underline {determination of $\eta$}
 $\to$ \underline {construction of the interaction}}.  
In the LO case one has $\eta^{(0)}(q)=q^2/a$, where
$a=\frac{g^2N}{2(2\pi)^{D-1}}$, and in general we shall assume at the NLO
the form $\eq=\eta^{(0)}(q)(1+\xq)$.

We briefly describe the calculations leading to the full interaction
at NLO for the octet state \cite{bravac}.
We first consider the expressions for the trajectory:
\beq
\omega^{(1)}(q)=-\bar{g}^2q^{2\epsilon}\frac{{\rm \Gamma}^2(\epsilon)}
{{\rm\Gamma}(2\epsilon)} \quad , \quad
\omega^{(2)}(q)=-\bar{g}^4q^{4\epsilon}F(\epsilon)
\eeq
where
\beq
\bar{g}^2=\frac{Ng^2{\rm \Gamma}(1-\epsilon)}{(4\pi)^{2+\epsilon}} \quad ,
\quad F(\epsilon)=\frac{A}{\epsilon^2}-
\frac{B}{\epsilon}-C-\frac{\pi^2}{3}A
\eeq
The coefficients $A,B,C$ contains a part proportional to $N_f$ (the number
of quark flavours) which therefore isolate the quark contribution to the
reggeized gluon trajectory $\omega_Q^{(2)}(q)$, proportional on
$F_Q(\epsilon)$. Thus the quark contribution to $\eta$ or to the $\xi$
functions is fixed looking only at $\omega_Q^{(2)}(q)$.
Let us write this $\xi_Q$. Using an integral representation for
$\omega_Q^{(2)}(q)$ one can write
\beq
2bq^2\int\frac{d^{D-2}q_1}{q_1^2q_2^2}(q^{2\epsilon}-
q_1^{2\epsilon}-q_2^{2\epsilon})=
-aq^2\int\frac{d^{D-2}q_1}{q_1^2q_2^2}[\xq-\xa-\xb]
\eeq
where the $\omega_Q^{(2)}(q)$ on the LHS is matched to the form (\ref{traj})
which solves the bootstrap equation and expanded to the second order; $b$ is
a function of the parameter $\epsilon$ with a pole in zero.

The last relation is solved for $ \xi_Q(q)=-2b q^{2\epsilon}/a$
and substituting this in the second order part of the potential (\ref{pot})
one easily obtains, for the quark contribution,
\[
V_Q^{(g,2)}(q_1,q'_1)=-b\frac{1}{\sqrt{q_1^2q_2^2{q'_1}^2{q'_2}^2}}
\Big[\frac{q_1^2{q'_2}^2}{k^2}
(q_1^{2\epsilon}+{q'_2}^{2\epsilon}-q_2^{2\epsilon}-{q'_1}^{2\epsilon}
-2k^{2\epsilon})+\]\beq
\frac{q_2^2{q'_1}^2}{k^2}
(q_2^{2\epsilon}+{q'_1}^{2\epsilon}-q_1^{2\epsilon}-{q'_2}^{2\epsilon}
-2k^{2\epsilon})
-q^2(2q^{2\epsilon}-q_1^{2\epsilon}-q_2^{2\epsilon}-{q'_1}^{2\epsilon}
-{q'_2}^{2\epsilon})
\Big]
\eeq
This expression (multiplied by $\sqrt{q_1^2q_2^2{q'_1}^2{q'_2}^2}$ to
match the integration measure) coincides with
the one found later in \cite{fafiopa} by direct methods (see Eq. (47) there).

We try to use this procedure to calculate the total gluon interaction, with also the
gluonic contribution. 
From the relations $\xi(q)/\xi_Q(q)=\omega^{(2)}(q)/\omega_Q^{(2)}(q)=
F(\epsilon)/F_Q(\epsilon)$ the answer is
\beq
V^{(g,2)}(q_1,q'_1)=V_Q^{(g,2)}(q_1,q'_1) F(\epsilon)/F_Q(\epsilon)
\eeq

Since the full kernel $K$, which solves the bootstrap relation, can be written
in an explicitely infrared finite way it is convenient to write the
expressions using the renormalized coupling constant $g_\mu$.
In the dimensional regularization in the $\overline{{\rm MS}}$ scheme one has
$g^2=g_{\mu}^2\mu^{-2\epsilon}\left(1+A\bar{g}_{\mu}^2/\epsilon\right)$
and $ \bar{g}_{\mu}^2=g_{\mu}^2N{\rm \Gamma}(1-\epsilon)/
(4\pi)^{2+\epsilon}$. This means that we can also
renormalize the $\eta$ or the $\xi$ functions.
One has a finite
$\xi_{\mu}(q)=\xi(q)-A\bar{g}_{\mu}^2/\epsilon$
because the poles in $1/\epsilon$ cancel. Therefore we are allowed to put
$\epsilon=0$ in the infrared finite expression for the non forward kernel in
the octet channel. The simple form we present here refers to a metric
modified in accord to $\Psi(q_1)\rightarrow\Psi(q_1)/\sqrt{\ea\eb}$.
For the second order contribution (as usual a superscript indicates the
order) we obtain
\[
K^{(2)}\Psi=\frac{3\alpha_s}{4\pi}\Bigl[ BK^{(1)}\Psi+
6A\alpha_s\int\frac{d^2q'_1q_1^2}{(2\pi)^2k^2}
\left(\frac{\Psi(q'_1)}{{q'_1}^2}-\frac{\Psi(q_1)}{{q'_1}^2+k^2}
\right)
\ln\frac{q_1^2\mu^2}{{q'_1}^2k^2}\]\beq+(1\leftrightarrow 2)
-6A\alpha_s\int\frac{d^2q'_1q^2\Psi(q'_1)}{(2\pi)^2{q'_1}^2{q'_2}^2}
\ln\frac{q^2\mu^2}{{q'_1}^2{q'_2}^2} \Bigr].
\eeq
Let us remark that
from some recent calculations \cite{bravac2} devoted to analyze
the NLO quark and gluon inpact factors calculated in \cite{ffkp}
we see that the ansatz given in (7-9) is correctly working for the
$q \bar{q}$ pair but not for the gluon contribution even if the
running of the coupling is correctly reproduced.
  
\section*{Acknowledgments}
The results shown have been obtained in collaboration with M. Braun
\cite{bravac}. The author is grateful for the support from the
Alexander von Humboldt Stiftung.

\end{document}